\documentclass[conference,a4paper]{IEEEtran} 
\IEEEoverridecommandlockouts
\usepackage[binary-units,per-mode=symbol,per-symbol=p]{siunitx}
\usepackage{cite}
\usepackage{adjustbox}
\usepackage[ruled,vlined]{algorithm2e}
\usepackage{amsmath,amssymb,amsfonts}
\usepackage{algorithmic}
\usepackage{graphicx}
\usepackage{textcomp}
\usepackage[colorlinks=false, pdfborder={0 0 0}, breaklinks]{hyperref}
\usepackage{xcolor}
\usepackage{balance}
\usepackage{enumitem}
\setlist{nolistsep,leftmargin=*}
\usepackage{setspace}
\usepackage{wrapfig}
\usepackage{listings}
\usepackage{color}
\usepackage{caption}
\usepackage{subcaption}
\usepackage{todonotes}
\usepackage[longtable]{multirow}
\usepackage{soul}

\def\BibTeX{{\rm B\kern-.05em{\sc i\kern-.025em b}\kern-.08em
    T\kern-.1667em\lower.7ex\hbox{E}\kern-.125emX}}

\usepackage{amsmath}

\usepackage{ctable}

\newcommand{\cf}{\emph{cf.}\xspace}

\newcommand{\eg}{\emph{e.g.}, }

\newcommand{\EY}{$E_{\text{Y}}$}

\newcommand{\LY}{$L_{\text{Y}}$}

\newcommand{\h}{$h$}

\newcommand{\BDRP}{BDR\textsubscript{P}}
\newcommand{\BDRV}{BDR\textsubscript{V}}
\newcommand{\BDRX}{BDR\textsubscript{X}}

\begin{document}
\bstctlcite{IEEEexample:BSTcontrol}
\title{Decoding Complexity-Rate-Quality Pareto-Front for Adaptive VVC Streaming}


\author{Angeliki Katsenou, Vignesh V Menon, Adam Wieckowski, Benjamin Bross, and Detlev Marpe
 
\thanks{Angeliki Katsenou is with Bristol Digital Futures Institute - BDFI and University of Bristol, UK. This work was partly funded by UKRI MyWorld Strength in Places Programme (SIPF00006/1).} 
\thanks{Vignesh V Menon, Adam Wieckowski, Benjamin Bross, and Detlev Marpe are with the Video Coding Systems research group at the Video Communication and Applications department, Fraunhofer HHI, Berlin.}
}

\maketitle

\begin{abstract}
Pareto-front optimization is crucial for addressing the multi-objective challenges in video streaming, enabling the identification of optimal trade-offs between conflicting goals such as bitrate, video quality, and decoding complexity. This paper explores the construction of efficient bitrate ladders for adaptive Versatile Video Coding (VVC) streaming, focusing on optimizing these trade-offs. We investigate various ladder construction methods based on Pareto-front optimization, including exhaustive Rate-Quality and fixed ladder approaches. We propose a joint decoding time-rate-quality Pareto-front, providing a comprehensive framework to balance bitrate, decoding time, and video quality in video streaming. This allows streaming services to tailor their encoding strategies to meet specific requirements, prioritizing low decoding latency, bandwidth efficiency, or a balanced approach, thus enhancing the overall user experience. The experimental results confirm and demonstrate these opportunities for navigating the decoding time-rate-quality space to support various use cases. For example, when prioritizing low decoding latency, the proposed method achieves decoding time reduction of \SI{14.86}{\percent} while providing Bjøntegaard delta rate savings of \SI{4.65}{\percent} and \SI{0.32}{\dB} improvement in the eXtended Peak Signal-to-Noise Ratio (XPSNR)-Rate domain over the traditional fixed ladder solution. 
\end{abstract}


\setlength{\textfloatsep}{1pt}

\section{Introduction}
\begin{figure}[t]
\centering
\begin{subfigure}{0.470\linewidth}
    \centering
    \includegraphics[width=0.98\textwidth]{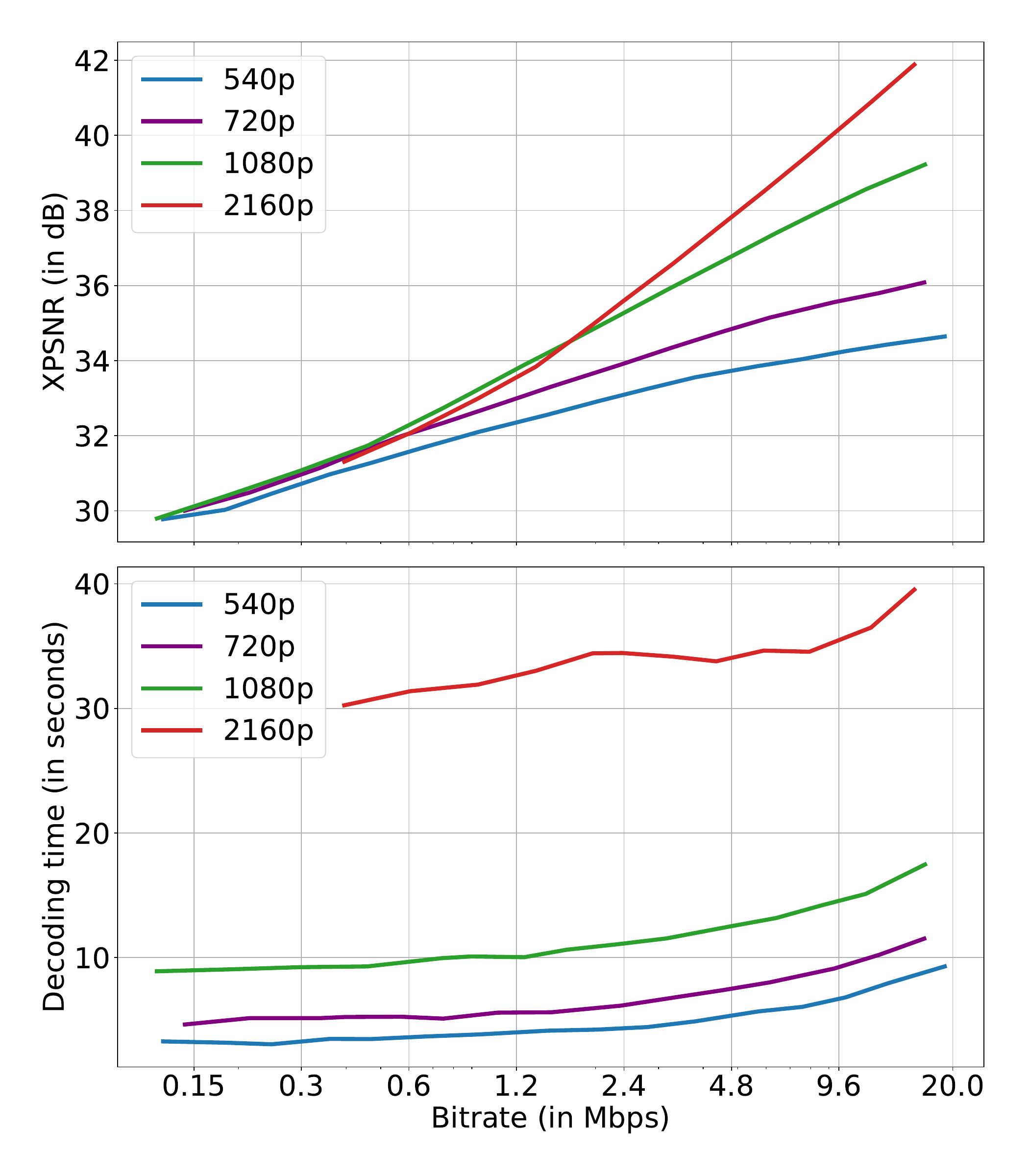}     
    \caption{Sequence \textit{0003}.}
\end{subfigure}
\hfill
\begin{subfigure}{0.470\linewidth}
    \centering
   \includegraphics[width=0.98\textwidth]{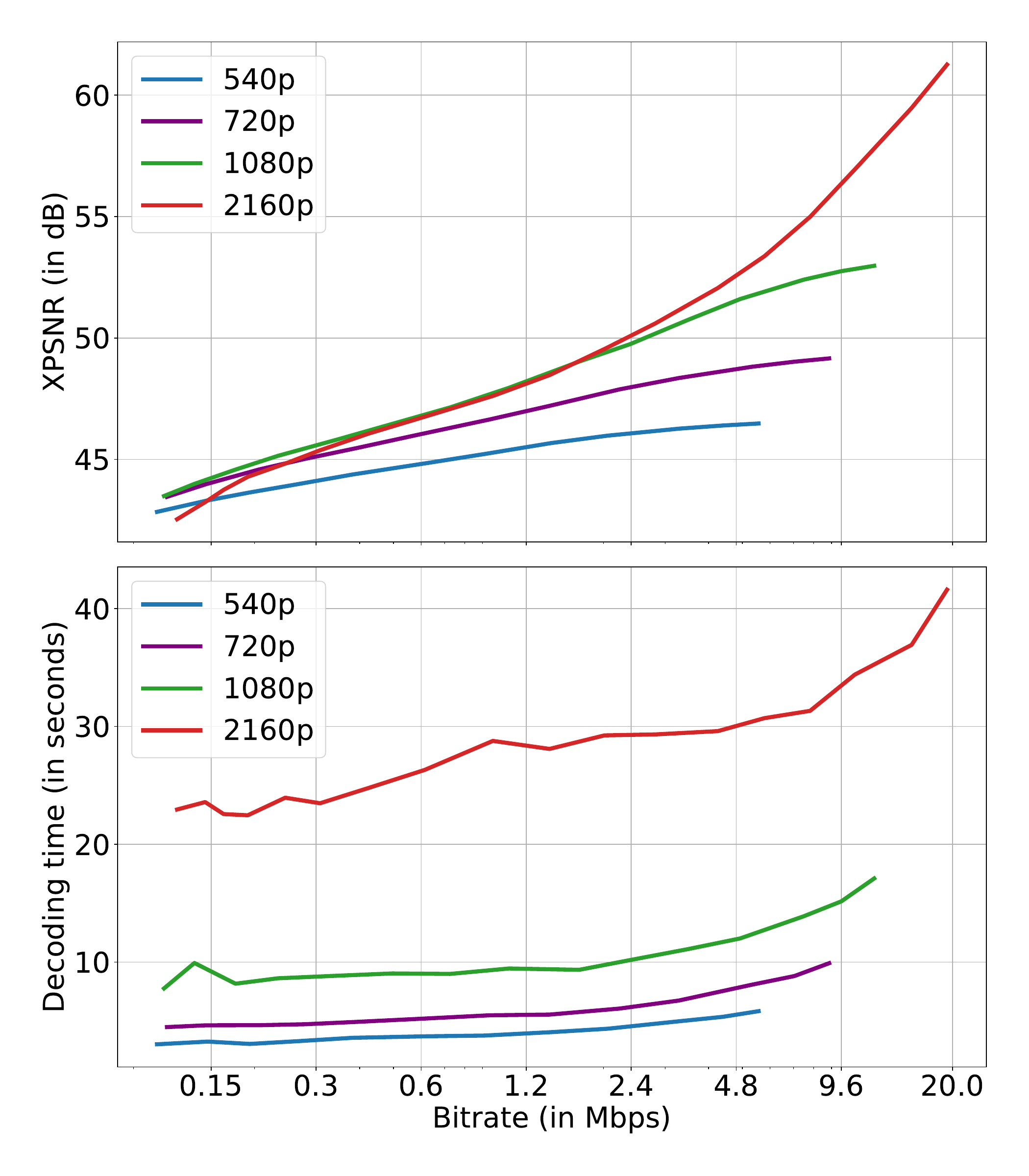}    
    \caption{Sequence \textit{0010}.}
\end{subfigure}
\caption{Rate-XPSNR curves and decoding times of example video sequences from the employed dataset, Inter4K~\cite{inter4k_ref}, encoded using VVenC~\cite{vvenc_ref} and decoded using VVdeC~\cite{VVdeC_ref}.}
\label{fig:intro_res}
\end{figure}

The increasing demand for high-quality video streaming over the internet has driven significant advancements in video compression technologies. Versatile Video Coding (VVC)~\cite{vvc_ref}, the latest video coding standard, promises substantial improvements in compression efficiency compared to its predecessors, including High Efficiency Video Coding (HEVC)~\cite{HEVC}. However, these advancements come with increased computational complexity, which poses challenges for energy-constrained client devices~\cite{vvc_complexity,hevc_vvc_enc_comp,dec_energy_vvc}. 
HTTP Adaptive Streaming (HAS) has become the \emph{de facto} standard for delivering online video content. HAS dynamically adjusts the quality of the video stream based on network conditions, user preferences, and device specifications, ensuring an optimal viewing experience across a wide range of conditions~\cite{HAS_survey}. To achieve this dynamic adjustment, video content is typically encoded at multiple bitrates and resolutions, creating a bitrate ladder from which the appropriate representation is selected in real-time per user~\cite{mpeg_dash_ref,farahani2024ai_sustainable}. Hence, constructing an efficient bitrate ladder is crucial for HAS, as it directly impacts both the Quality of Experience (QoE) of the user and the computational resources required for decoding, as illustrated in Fig.~\ref{fig:intro_res}. 

Traditional ladder construction methods often rely on fixed~\cite{HLS_ladder_ref} or heuristic-based approaches, primarily focused on bitrate-quality curves that do not adequately balance the trade-offs between bitrate, video quality, and decoding complexity. Pareto-front (PF) optimization offers a robust framework for addressing these trade-offs. By identifying a set of optimal solutions representing the best possible compromises between conflicting objectives, PF optimization techniques enable the construction of bitrate ladders that maximize video quality while minimizing bitrate and decoding complexity. Several quality metrics can be employed in this context, such as eXtended Peak Signal-to-Noise Ratio (XPSNR)~\cite{XPSNR} and Video Multi-Method Assessment Fusion (VMAF)~\cite{VMAF,Katsenou_vmaf_ladder}, each providing a different perspective on video quality. However, it is observed that PSNR and VMAF measures fail to model the subjective quality of VVC-coded ultra high definition (UHD) bitstreams accurately~\cite{xpsnr_vs_vmaf}. Moreover, it was observed that XPSNR can predict the subjective codec ranking reported in~\cite{wien_xpsnr_vs_vmaf} with acceptable accuracy~\cite{itu_xpsnr_vs_vmaf}. 
Furthermore, reducing decoding runtime correlates with lower energy consumption, which is crucial for battery-powered devices~\cite{dec_time_energy_rel}. PF optimization contributes to energy-efficient video playback, extending battery life and reducing the environmental impact. Moreover, most consumers accept lower video quality (to a certain extent) if they can consume less energy and save money~\cite{qoe_energy_tradeoff}. This paper explores developing and evaluating various ladder construction methods based on PF optimization, specifically targeting VVC streaming.

This work evaluates these methods to provide insights into their effectiveness in real-world adaptive streaming scenarios. The main objective is to advance the understanding of complexity-quality trade-offs in adaptive VVC streaming that can contribute to developing more efficient and effective streaming solutions. Our contributions include a detailed analysis of each method's trade-offs between bitrate, video quality, and decoding runtime and practical recommendations for their implementation in VVC streaming systems. In the following sections, we discuss related work in Section~\ref{sec: SotA_PFStreaming}, describe our methodology in Section~\ref{sec: Methodology}, present the experimental setup in Section~\ref{sec: Evaluation}, and discuss the results in Section~\ref{sec: Results}. Finally, the paper concludes with Section~\ref{sec: Conclusions}.

\section{Pareto-front optimization in streaming}
\label{sec: SotA_PFStreaming}
PF optimization is a multi-objective optimization technique to identify optimal solutions representing the best possible trade-offs between different objectives. In video streaming, PF optimization helps balance bitrate, video quality, and decoding complexity, allowing the selection of the most efficient encoding configurations to meet specific requirements~\cite{Katsenou_IEEEOJSP2021}. 
Previous research efforts~\cite{Katsenou_IEEEOJSP2021, Katsenou_vmaf_ladder, ladre_ref} have explored various aspects of PF optimization for video streaming. For instance, some studies have investigated rate-quality PF to minimize bitrate while maintaining acceptable video quality~\cite{netflix_paper, Katsenou_vmaf_ladder}. Others have focused on complexity-quality trade-offs, aiming to reduce computational complexity without significant loss in video quality~\cite{ladre_ref}. These approaches lay the foundation for more sophisticated bitrate ladder construction methods.

Traditional methods construct bitrate ladders, aka defining streaming parameters such as resolution and compression level, often relying on fixed~\cite{HLS_ladder_ref} or heuristic-based approaches~\cite{KokaramICIP2018}, which may not fully exploit the potential of PF optimization. Recent advancements have proposed more dynamic and adaptive techniques, incorporating various quality metrics and computational considerations~\cite{jtps_ref, ensemble_learning_vvc_ladder, ML_PTE_survey}. Another significant contribution is the integration of decoding complexity into the optimization process, addressing the need for efficient playback on resource-constrained devices~\cite{Herglotz_2019}. Finally, an exploration of the potential optimization in the energy-quality space instead of the rate-quality has been featured in~\cite{Katsenou_ICIP2024}, revealing opportunities for sustainable streaming.

This paper builds on these foundations by exploring the application of PF optimization to construct bitrate ladders that balance bitrate, video quality, and decoding complexity in the context of VVC streaming. 

\section{Joint Rate-Quality-Decoding Time Pareto-front (RQT-PF)}
\label{sec: Methodology}
\begin{figure}[t]
\centering
\begin{subfigure}{0.489\linewidth}
    \centering
    \includegraphics[width=\textwidth]{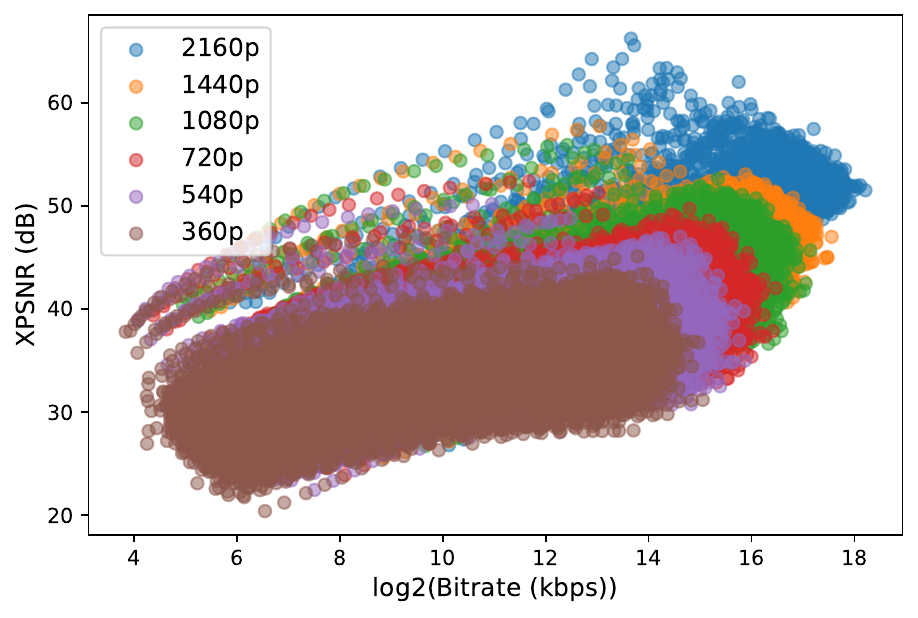}
    \caption{ RQ parameter space.}
\end{subfigure}
\begin{subfigure}{0.489\linewidth}
    \centering
    \includegraphics[width=\textwidth]{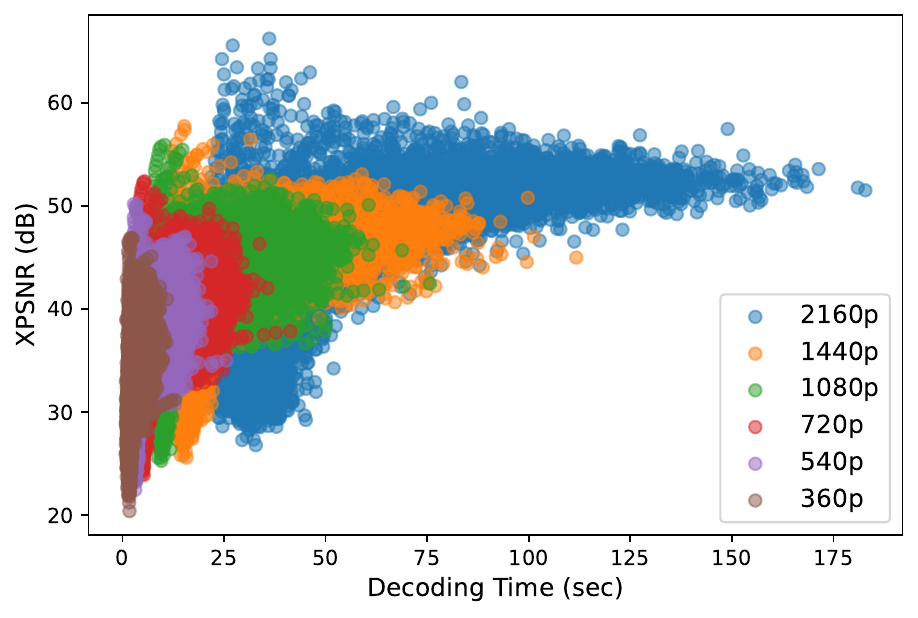}
    \caption{ TQ parameter space.}
\end{subfigure}
\caption{Quality-Rate-Time points for VVenC encodes across six spatial resolutions. }
\label{fig: ParamSpace}
\end{figure}

Adaptive streaming aims to deliver high-quality while minimizing playback interruptions caused by buffering. This requires carefully selecting encoding parameters, particularly resolution and quantization parameters (QP), to manage the trade-off between video quality and decoding complexity. Fig.~\ref{fig: ParamSpace} illustrates the Rate-Quality-Decoding Time parameter space across different quantization levels (QP) and spatial resolutions for the Inter4K dataset~\cite{inter4k_ref}. Higher resolutions and lower QP values result in better video quality at the cost of increased bitrate and decoding time. Conversely, lower resolutions and higher QP values reduce decoding time and bitrate but may compromise the delivered video quality. 
The apparent clustering of points for different resolutions in both subfigures highlights the distinct trade-offs between bitrate, quality, and decoding time for each resolution. Analyzing these clusters can identify optimal encoding parameters that balance quality, bitrate, and decoding time. The visual separation of resolutions suggests that adaptive streaming solutions can tailor their bitrate ladders more precisely based on the desired balance between quality and decoding complexity.

Algorithm~\ref{algo:rqt_pf} illustrates the proposed \emph{RQT-PF} estimation. 
This three-dimensional PF is crucial for adaptive streaming scenarios where bandwidth efficiency and computational complexity must be optimized simultaneously. By identifying the optimal points on this PF, encoding schemes can be developed that deliver high-quality video at efficient bitrates and with acceptable decoding times.
Towards this realization, we define a composite metric $M$ as a linear combination of decoding time and bitrate, weighted by parameter $\alpha$:
\begin{align}
    M = \alpha \cdot log (\tau_{D}) + (1-\alpha) \cdot log (b),
    \label{eq: M}
\end{align}
where $\alpha\in \mathbb{R}$.
The logarithmic transformation is applied to decoding time and bitrate to appropriately scale their wide range of values. The goal is to minimize the composite metric $M$ while maximizing the video quality $v$. This can be formulated as a multi-objective optimization problem:


\begin{equation*}
        \min_{r, q} M = \min_{r, q} (\alpha \log_{10}(\tau_{\text{D}}(r,q)) + (1-\alpha) \log_{10}(b(r,q))) 
\end{equation*}
\vspace{-.5em}
\begin{equation*}
        \max_{r, q} v (r,q)
\end{equation*}
\vspace{-.5em}
\begin{equation}
    \begin{aligned}
\text{subject to}& \; \alpha \in [0, 1] \, .&\\
    \end{aligned}
\label{eq: Mopt}
\end{equation}

\begin{algorithm}[t]
\caption{\emph{RQT-PF} Estimation}
\small
\textbf{Input:}\\
\quad $\mathcal{R}$~: set of supported resolutions\\
\quad $\mathcal{Q}$~: set of quantization parameters (QP)\\
\begin{algorithmic}[1]
\STATE Load video content \;
\FOR{each $r \in \mathcal{R}$}
    \FOR{each $q \in \mathcal{Q}$}
        \STATE Encode video using ($r, q$) \;
        \STATE Compute the bitrate ($b$) \;
        \STATE Measure decoding time ($\tau_{D}$) \;
        \STATE Calculate quality $v$  \; 
    \ENDFOR
\ENDFOR
\STATE Perform multi-objective optimization in the $M-v$ (RQT) space to identify PF points as in Eq.(\ref{eq: M}-\ref{eq: Mopt})\;
\STATE Sample the PF to construct the bitrate ladder \;
\end{algorithmic}
\label{algo:rqt_pf}
\end{algorithm}

The PF consists of a set of encoding configurations for which it is impossible to improve one objective without degrading the other. Mathematically, a point ($\tau_{D}, b, v$) is on the PF, if and only if there is no other point ($\tau_{D}', b', v'$) such that:
\begin{align}
(\tau_{D}' \leq \tau_{D}) \land (b' \leq b) \land (v' \geq v)
\end{align}
with at least one strict inequality.

Streaming services can use these PFs to select the most appropriate encoding configurations based on their constraints and user requirements. For instance, a service prioritizing low latency would choose configurations from the PF with $\alpha = 0.75$, while another focusing on bandwidth savings would prefer $\alpha = 0.25$. A value of $\alpha = 1$ would mean that the priority is to minimize the decoding time over bitrate while ensuring the best quality. Thus, the PF would be based only on Quality-Decoding Time Curves across spatial resolutions. We will refer to this method as \emph{QT-PF}. Finally, a value of $\alpha = 0$ would exclude decoding time, and the PF constructed would be exclusively based on the Rate-Quality curves as standard practice (\eg \cite{netflix_paper}). We will henceforth refer to this method as \emph{DynResXPSNR} introduced in~\cite{menon2024convexhull_xpsnr}.

\section{Evaluation Setup}
\label{sec: Evaluation}

\subsection{Benchmarks}
The proposed method is compared against the following state-of-the-art methods:
\begin{enumerate}
    \item \emph{FixedLadder}~\cite{HLS_ladder_ref}: We use the HLS bitrate ladder specified by Apple as the fixed set of bitrate-resolution pairs. 
    \item \emph{Default}: We only encode at the native, 2160p, resolution for a given set of bitrates. This method demonstrates a limitation in covering the whole range of the bitrate ladder, especially the low bitrates. 
    \item \emph{DynResXPSNR}~\cite{menon2024convexhull_xpsnr}: The resolution yielding the highest XPSNR is selected for a given set of bitrates. 
\end{enumerate}

\subsection{Dataset, codecs, and metrics}
Inter4K~\cite{inter4k_ref} comprises 1,000 UHD clips at 60 frames per second (fps), sourced from YouTube. The dataset includes standardized video resolutions ranging from ultra-high definition (UHD/4K) and quad-high definition (QHD/2K) to full-high definition (FHD/1080p), high definition (HD/720p), quarter-high definition (qHD/520p), and ninth-high definition (nHD/360p). Inter4K is organized into six main categories based on the primary focus of each video: urban environments (\eg buildings, streets, or vehicles),
nature and animals, sports and people (depicting human activities and actions),
demos and abstracts (including demo videos for video resolution and frame rates, or videos with computer-generated abstract shapes), music videos, and movies.

\begin{table}[t]
\caption{Experimental parameters and values.}
\centering
\resizebox{0.95\columnwidth}{!}{
\begin{tabular}{l||c|c|c|c|c|c}
\specialrule{.12em}{.05em}{.05em}
\specialrule{.12em}{.05em}{.05em}
\emph{Parameter} & \multicolumn{6}{c}{\emph{Values}}\\
\specialrule{.12em}{.05em}{.05em}
\specialrule{.12em}{.05em}{.05em}
$\mathcal{R}$ & \multicolumn{6}{c}{\{ 360, 540, 720, 1080, 1440, 2160 \} } \\
\hline
$\mathcal{B}$ & 0.145 & 0.300 & 0.600 & 0.900 & 1.600 & 2.400 \\
\cmidrule{2-7}
              & 3.400 & 4.500 & 5.800 & 8.100 & 11.600 & 16.800 \\
\hline
$\mathcal{Q}$ &  \multicolumn{6}{c}{10, 12, \ldots, 50} \\
\hline
$\alpha$ & \multicolumn{2}{c|}{0.25} & \multicolumn{2}{c|}{0.5} & \multicolumn{2}{c}{0.75}\\
\hline
Target encoder & \multicolumn{6}{c}{VVenC [faster][intra period=\SI{1}{\second}][4 CPU threads]} \\
\hline
Target decoder & \multicolumn{6}{c}{VVdeC [4 CPU threads]} \\
\specialrule{.12em}{.05em}{.05em}
\specialrule{.12em}{.05em}{.05em}
\end{tabular}
}
\label{tab:exp_par}
\end{table}

The experimental parameters used in this paper are reported in Table~\ref{tab:exp_par}. We run all experiments on a dual-processor server with Intel Xeon Gold 5218R (80 cores, frequency at \SI{2.10}{\giga\hertz}), where each VVenC and VVdeC instance uses four CPU threads. 
We compare the overall quality in PSNR~\cite{psnr_ref}, VMAF~\cite{VMAF}, and XPSNR~\cite{XPSNR} and the achieved bitrate for every target bitrate of each test sequence. Bjøntegaard Delta metrics~\cite{BJDelta} are computed over Rate-PSNR, Rate-XPSNR, and Rate-VMAF ladders. Particularly, rates \BDRP, \BDRX, and \BDRV~refer to the average increase in bitrate of the considered methods compared with the \emph{FixedLadder} encoding with the same \mbox{PSNR}, \mbox{XPSNR} and \mbox{VMAF}, respectively.

Additionally, we compute the decoding time difference, $\Delta T_{\text{D}}$, as an important performance indicator of an adaptive video streaming solution:
\begin{align}
 \Delta T_{\text{D}} &= \frac{\sum \tau_{method} - \sum \tau_{ref}}{\sum \tau_{ref}}
\end{align}
where $method$ represents the evaluated method and $ref$ the reference method, which in our case is the \textit{FixedLadder}.

\section{Experimental results}
\label{sec: Results}

\begin{figure}[t]
\centering
\begin{subfigure}{0.493\linewidth}
    \centering
    \includegraphics[width=\textwidth]{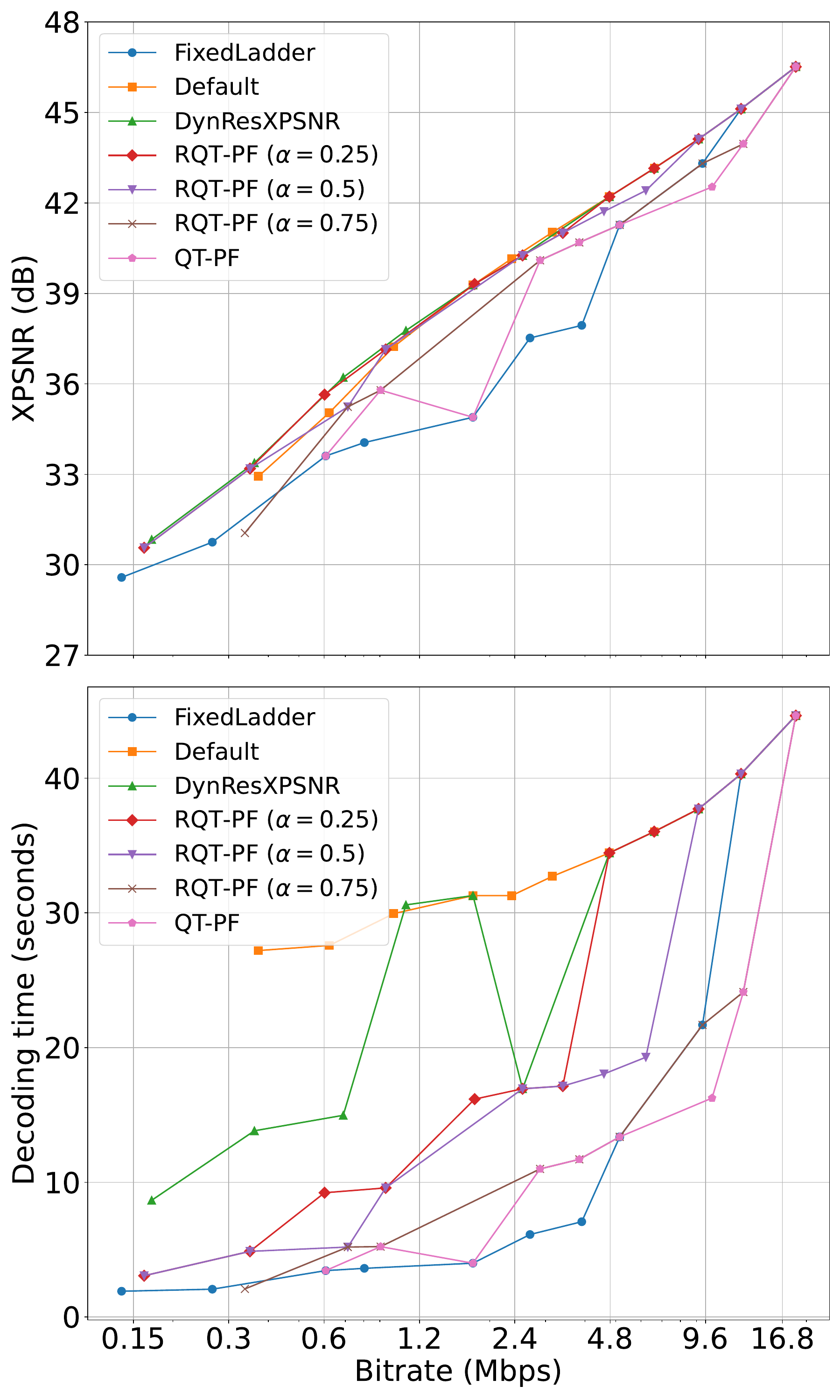} 
    \caption{Sequence \textit{0153}\\ (\EY:50.83, \h:9.80, \LY:93.54)}
\end{subfigure}
\begin{subfigure}{0.493\linewidth}
    \centering
   \includegraphics[width=\textwidth]{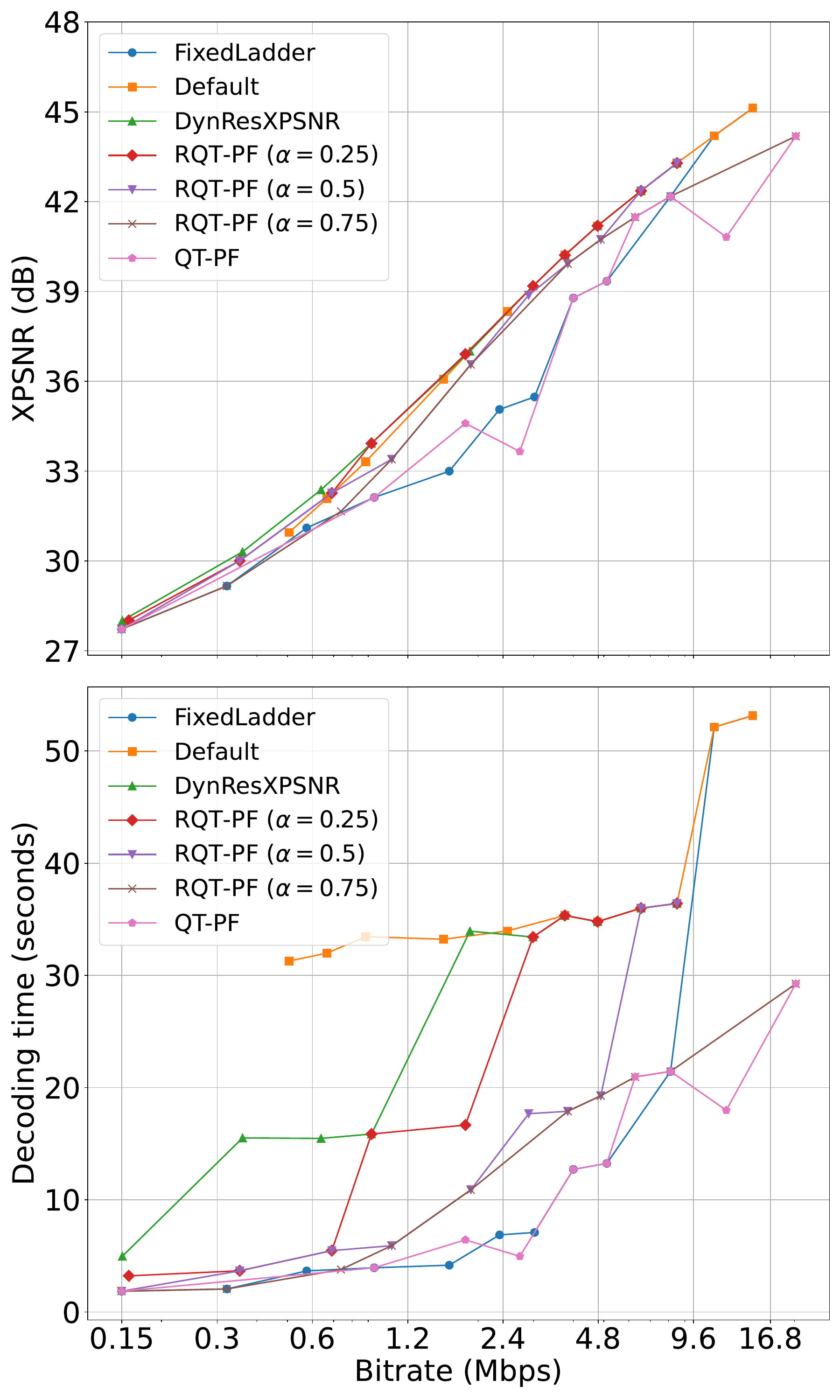}    
    \caption{Sequence \textit{0263}\\ (\EY:122.64, \h:38.48, \LY:116.31)}
\end{subfigure}

\caption{Resulting ladders of example video sequences. 
}
\label{fig:rd_res}
\end{figure}

A visual inspection of the ladders produced by \emph{RQT-PF} and the benchmarks is feasible through the examples of Fig.~\ref{fig:rd_res}. Two sequences were selected based on their different spatio-temporal features extracted using the Video Complexity Analyzer (VCA) tool~\cite{vca_ref}. The scene in sequence \emph{0153} is a night capture of part of the Eiffel Tower with a panning camera, resulting in moderate values on all three reported features, texture energy (\EY), temporal texture energy (\h), and average luminance (\LY). The scene in sequence \emph{0263} has different content features: high spatial texture energy (\EY) and temporal texture energy (\h) but moderate average luminance (\LY). It depicts a moving crowd of tourists in front of a sculpted building with reflections on its window muntins, resulting in. The constructed ladders follow similar patterns for these two sequences. Significant interlacing exists across the different solutions in the XPSNR-Rate plots, mainly due to the different resolution switches across the ladder rungs. The \emph{RQT-PF} with $\alpha=0.25$ is tightly close to \emph{DynResXPSNR} along the whole range in the top figures indicating the best rate-quality performance. As anticipated, the \emph{Default} curve meets the quality-optimal solutions at high bitrates, where transmitting the native resolution provides the best quality-rate tradeoff. Moving to the bottom plots, we notice that the \emph{Default} and \emph{DynResXPSNR} result in the highest decoding times for most rungs.

\begin{table}[tb]
\caption{Average performance of the explored methods and benchmarks compared against \emph{FixedLadder}. The best average performance per column is annotated with \textbf{bold}, and the second best is \underline{underlined}.}
\centering
\resizebox{\columnwidth}{!}{
\begin{tabular}{@{}l@{ }||@{ }c@{ }|@{ }c@{ }|@{ }c@{ }|@{ }c@{ }|@{ }c@{ }|@{ }c@{ }|@{ }c@{ }}
\specialrule{.12em}{.05em}{.05em}
\specialrule{.12em}{.05em}{.05em}
Method & \BDRP & \BDRX & \BDRV & BD-PSNR & BD-XPSNR & BD-VMAF  &  $\Delta T_{\text{D}}$\\
& [\%] & [\%] & [\%] & [dB] & [dB] & & [\%] \\
\specialrule{.12em}{.05em}{.05em}
\specialrule{.12em}{.05em}{.05em}
\emph{Default} & -25.53 & -18.46 & \underline{-42.41} & \underline{1.35} & 1.01 & \underline{8.49} & 159.39 \\
\hline
\emph{DynResXPSNR} & \textbf{-28.97} &	\textbf{-24.96} & \textbf{-46.43} & \textbf{1.34} & \textbf{1.11} & \textbf{9.79} & 72.68 \\
\hline
\emph{QT-PF} & 13.45 & 12.73 & 37.59 & -0.25 & -0.24 & -2.71  & \textbf{-27.74}\\
\hline
\emph{RQT-PF ($\alpha=0.75$)} & -4.60 & -4.65 & 3.80 & 0.34 & 0.32 & 0.71 & \underline{-14.86}\\
\emph{RQT-PF ($\alpha=0.5$)} & -18.00 & -17.55 & -18.82 & 0.85 & 0.80 & 4.37 & 3.98 \\
\emph{RQT-PF ($\alpha=0.25$)} & \underline{-26.25} & \underline{-24.74} & -37.40 & 1.21 & \underline{1.10} & 7.69 & 34.37 \\
\specialrule{.12em}{.05em}{.05em}
\specialrule{.12em}{.05em}{.05em}
\end{tabular}}
\label{tab:res_cons_energy}
\end{table}

Further results in Table~\ref{tab:res_cons_energy} compare the average encoding performance of various encoding methods against \textit{FixedLadder} encoding. The \emph{Default} method shows significant bitrate savings for the range it can cover, with \BDRP, \BDRX, and \BDRV~values of \SI{-25.53}{\percent}, \SI{-18.46}{\percent}, and \SI{-42.41}{\percent}, respectively, but with a significant increase of \SI{159.39}{\percent} in decoding time. \emph{DynResXPSNR} performs even better in terms of bitrate reduction and quality improvement, with the highest reductions in bitrate (\BDRP: \SI{-28.97}{\percent}, \BDRX: \SI{-24.96}{\percent}, \BDRV: \SI{-46.43}{\percent}). However, this comes at the cost of a substantial increase of \SI{72.68}{\percent} in decoding time. The focus of \emph{QT-PF} on quality and decoding time shows a decrease in decoding time, with a $\Delta T_{\text{D}}$ of \SI{-27.74}{\percent}, indicating the opportunity for high gains in reducing decoding time. The \emph{RQT-PF} methods with different $\alpha$ values operate between the \emph{DynResXPSNR} and \mbox{\emph{QT-PF}}. For $\alpha=0.75$, \emph{RQT-PF} achieves moderate bitrate reduction (\BDRP: \SI{-4.60}{\percent}, \BDRX: \SI{-4.65}{\percent}, but with increase for \BDRV: \SI{3.80}{\percent}) with a significant decrease in decoding time by \SI{14.86}{\percent}, indicating a strong effectiveness on decoding complexity reduction. As $\alpha$ decreases to $0.5$ and $0.25$, \emph{RQT-PF} shows improved bitrate savings and better quality, but with varying impacts on decoding time. These results highlight the flexibility of \emph{RQT-PF} in balancing bitrate, quality, and decoding time according to the chosen $\alpha$ value.

\begin{figure}[t]
\centering
\begin{subfigure}{0.49\linewidth}
    \centering
    \includegraphics[width=\textwidth]{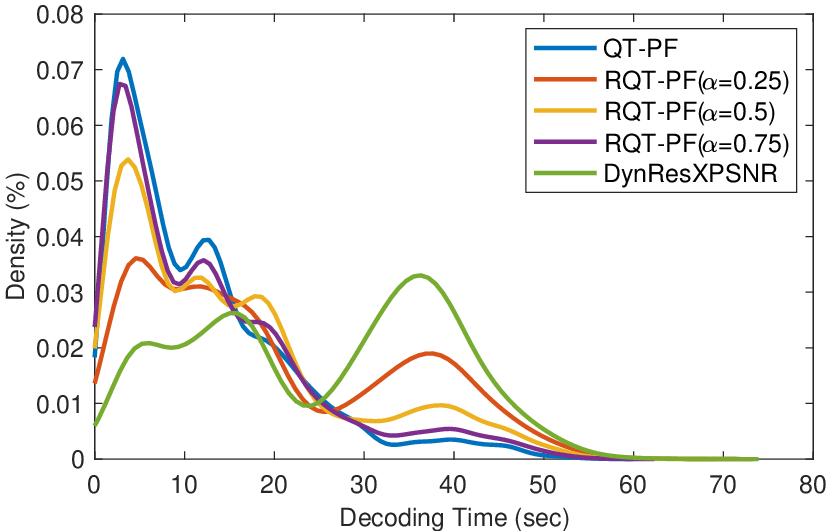}
    \caption{Decoding Time.}
    \label{fig:pdf_dec_time}
\end{subfigure}
\begin{subfigure}{0.49\linewidth}
    \centering
    \includegraphics[width=1.05\textwidth]{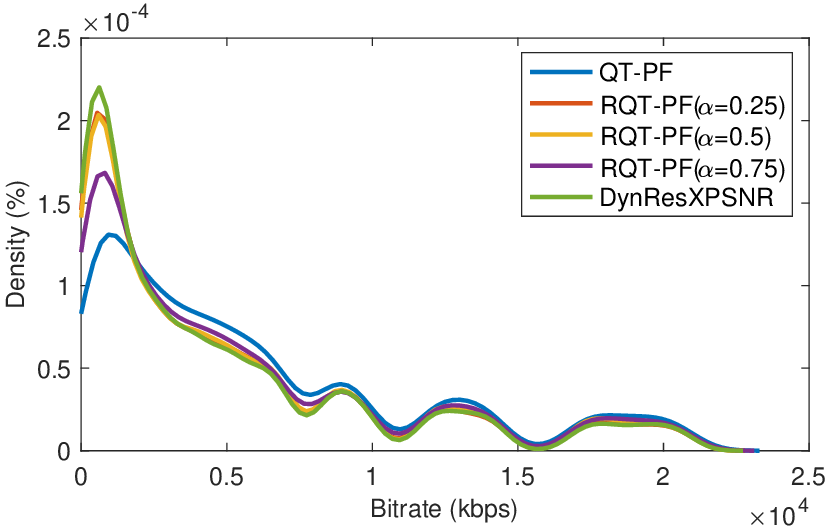}
    \caption{Bitrate.}
    \label{fig:pdf_bitrate}
\end{subfigure}
\begin{subfigure}{0.49\linewidth}
    \centering
    \includegraphics[width=\textwidth]{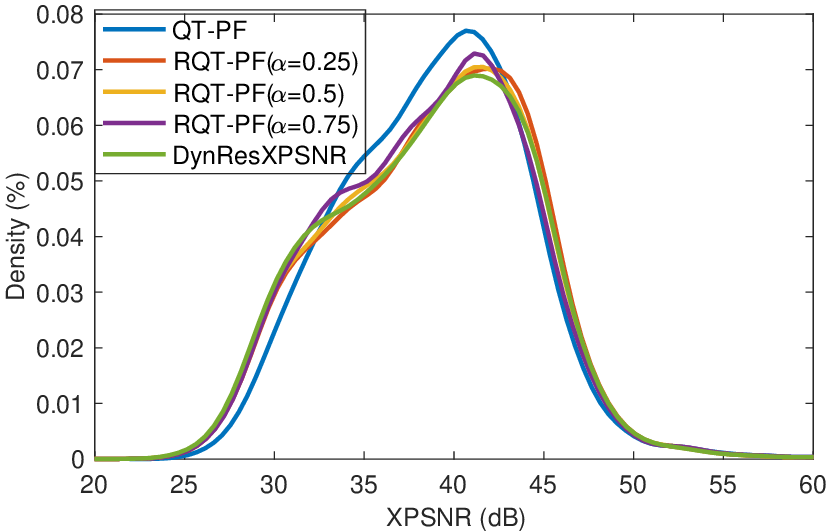}
    \caption{XPSNR.}
    \label{fig:pdf_xpsnr}
\end{subfigure}
\begin{subfigure}{0.49\linewidth}
    \centering
    \includegraphics[width=1.01\textwidth]{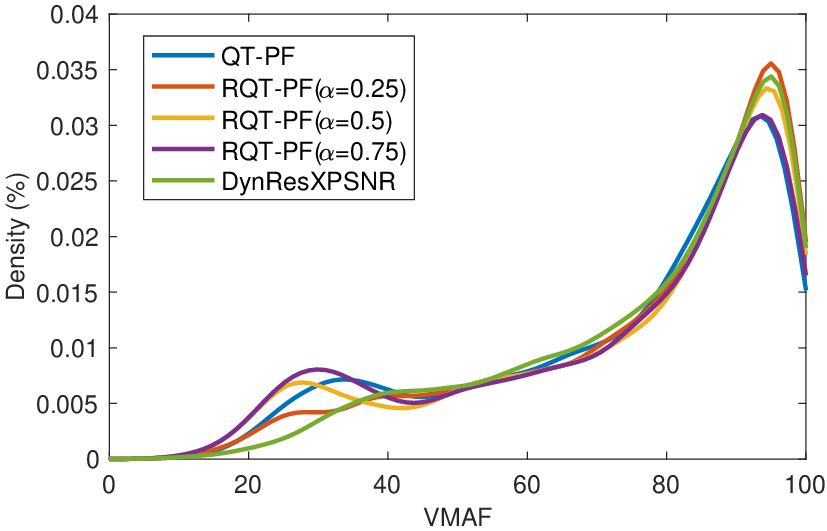}
    \caption{VMAF.}
    \label{fig:pdf_vmaf}
\end{subfigure}
\caption{Decoding time, bitrate, and quality PDFs for the \emph{RQT-PF}, \textit{QT-PF}, and \textit{DynResXPSNR} Ladders.}
    \label{fig: PDFs}
\end{figure}

It is evident that the lower the $\alpha$ value, the more similar the ladder to \emph{DynResXPSNR} is. The higher the $\alpha$ value, the more similar to the \emph{QT-PF} ladder we obtain. These shifting tradeoffs are confirmed by the probability density functions (PDFs) of Fig.~\ref{fig: PDFs}. The decoding time PDFs highlight the notable difference between the \emph{QT-PF} (blue line) and the \emph{DynResXPSNR} (green line) methods, with the \emph{QT-PF} approach significantly skewing and peaking towards lower decoding times. 
On the other hand, the bitrate distributions remain relatively consistent across methods (\cf Fig.~\ref{fig:pdf_bitrate}). \emph{RQT-PF}, particularly with lower $\alpha$ values, maintain a competitive XPSNR distribution (\cf Fig.~\ref{fig:pdf_xpsnr}). Finally, the VMAF PDFs in Fig.~\ref{fig:pdf_vmaf} indicate that the \emph{DynResXPSNR} (green line) and \emph{RQT-PF} $\alpha=\{0.25,0.5\}$ (red and yellow lines) methods yield higher VMAF values. Overall, these PDFs illustrate the effectiveness of the \emph{RQT-PF} methods in balancing bitrate, decoding time, and quality metrics, demonstrating their potential to support streaming use cases that require operation at various tradeoffs.

\section{Conclusions}
\label{sec: Conclusions}
This paper explored the application of PF-based bitrate ladder construction for adaptive VVC streaming, balancing the trade-offs between bitrate, video quality, and decoding runtime to enhance the efficiency of video streaming on energy-constrained devices. Our proposed \emph{RQT-PF} methodology was compared against content-gnostic and agnostic state-of-the-art approaches. Our experimental results demonstrated that \emph{RQT-PF} could effectively balance the competing objectives of bitrate, quality, and decoding runtime. By incorporating decoding time into the optimization process, \emph{RQT-PF} significantly reduced decoding complexity without substantial loss in video quality. This makes it particularly suitable for scenarios involving battery-powered devices and energy-efficient streaming. Future work intends to investigate whether time can serve as a proxy for energy optimization and perform a subjective quality of the proposed solutions.
\balance
\bibliographystyle{IEEEtran}
\bibliography{references.bib}
\balance
\end{document}